# Annealing induced colossal magnetocapacitance and colossal magnetoresistance in In-doped $CdCr_2S_4$

Y. M. Xie,[1] Z. R. Yang,[1,a)] L. Li,[1] L. H. Yin,[1] X. B. Hu,[1] Y. L. Huang,[1] H. B. Jian,[1] W. H. Song,[1] Y. P. Sun,[1] S. Q. Zhou,[2] and Y. H. Zhang[3]

[1]Key Laboratory of Materials Physics, Institute of Solid State Physics, Chinese Academy of Sciences, Hefei 230031, People's Republic of China

[2]Institute of Ion Beam Physics and Materials Research, Helmholtz-Zentrum Dresden-Rossendorf, P. O. Box 510119, Dresden 01314, Germany

[3]High Magnetic Field Laboratory, Hefei Institutes of Physical Science, Chinese Academy of Sciences, Hefei 230031, People's Republic of China

The correlation between colossal magnetocapacitance (CMC) and colossal magnetoresistance (CMR) in $CdCr_2S_4$ system has been revealed. The CMC is induced in polycrystalline $Cd_{0.97}In_{0.03}Cr_2S_4$ by annealing in cadmium vapor. At the same time, an insulator-metal transition and a concomitant CMR are observed near the Curie temperature. In contrast, after the same annealing treatment, $CdCr_2S_4$ displays a typical semiconductor behavior and does not show magnetic field dependent dielectric and electric transport properties. The simultaneous occurrence or absence of CMC and CMR effects implies that the CMC in the annealed $Cd_{0.97}In_{0.03}Cr_2S_4$ could be explained qualitatively by a combination of CMR and Maxwell-Wagner effect.

---

a)Corresponding author. Email address: zryang@issp.ac.cn

Multiferroic materials that exhibit simultaneous magnetic and ferroelectric order as well as concomitant magnetoelectric coupling have attracted special interest in recent years.[1-3] $CdCr_2S_4$ was originally investigated as a ferromagnetic semiconductor more than 40 years ago.[4] Recently it was reported that $CdCr_2S_4$ single crystal exhibits multiferroic behavior with the evidence of relaxor ferroelectricity and colossal magnetocapacitance (CMC).[5] Soon after, similar magnetoelectric effect has also been revealed in other related thio-spinel compounds, i.e. $CdCr_2Se_4$ and $HgCr_2S_4$.[6,7] However, the emergence of ferroelectricity and CMC effect in these thio-spinels was found to be highly sensitive to the detail of sample preparation and chemical doping. Annealing single crystal samples in vacuum or sulphur atmosphere led to a suppression of relaxation features and no remanent electric polarization could be found at low temperatures.[8] Moreover, multiferroicity and CMC effect are absent in undoped polycrystalline samples, but present in indium doped polycrystallines.[8] Accordingly, whether the magnetoelectric effect in $CdCr_2S_4$ is intrinsic or not has been raised.[9-11] Based on first principles calculations, Fennie and Rabe suggested that the anomalous dielectric response in $CdCr_2S_4$ might be extrinsic since this material shows no phonon softening at any temperature.[9] Catalan and Scott argued that the CMC effect and relaxor ferroelectricity reported in the thio-spinels could be correlated to Maxwell-Wagner effect caused by chlorine-based impurity or sulfur deficiency.[10,11] Nevertheless Hemberger et al. replied that they had excluded any nonhomogeneous impurity distribution by electron probe microanalysis and X-ray studies.[12] More recently, field-dependent dielectric and magnetic properties of $CdCr_2S_4$ single crystals have been further studied by Sun et al., and they suggested that the exchange striction might be responsible for the magnetoelectric phenomena in $CdCr_2S_4$.[13,14]

In this letter, we attempt to understand the origin of the CMC effect in the $CdCr_2S_4$ system. We investigate the magnetic, dielectric and electric transport properties of $CdCr_2S_4$ and $Cd_{0.97}In_{0.03}Cr_2S_4$ polycrystalline samples before and after annealing in cadmium vapor. We show that annealing has marginal effect on the magnetism, but greatly changes the magneto-transport and magneto-dielectric behaviors. After annealing $Cd_{0.97}In_{0.03}Cr_2S_4$, its temperature dependence of resistivity displays a metal-insulator transition around the Curie temperature, moreover both CMC and colossal magnetoresistance (CMR) are induced. In contrast, after the same annealing treatment, $CdCr_2S_4$ displays a typical semiconductor behavior and does not show magnetic field dependent dielectric and electric transport properties.

$CdCr_2S_4$ and $Cd_{0.97}In_{0.03}Cr_2S_4$ polycrystalline samples were prepared by a standard solid-state synthesis method.[15] The annealing treatment was performed by heating the sample pellet together with some cadmium metal particle in the quartz tube. $Cd_{0.97}In_{0.03}Cr_2S_4$ samples were annealed at 380 °C or 800 °C, while the $CdCr_2S_4$ sample was annealed only at 800 °C. The structure of samples was characterized by x-ray power diffraction (XRD) measurement. The magnetic properties were measured using a superconducting quantum interference device (MPMSXL-7) magnetometer. The resistivity of $Cd_{0.97}In_{0.03}Cr_2S_4$ sample annealed at 380 °C was measured using a two-probe method in Keithley 2182 Digital Sensitive VoltMeter and Keithley 220 Programmable Current Source integrated to physical properties measurement system (PPMS). Resistivity of the samples annealed at 800 °C was measured using a standard four-probe method in PPMS. Dielectric measurements were performed using an LCR meter (TH2828S) integrated to MPMS. Electric field was applied perpendicularly to the magnetic field.

The XRD analysis reveals that all the samples are single phase with spinel structure. Upon doping and annealing, no impurity phases are detected. The M - T curves under an external magnetic field of 1 T for all samples are shown in the inset of Fig. 1(b) and Fig. 2(b). It can be found that, after annealing in cadmium vapor, the magnetic properties of $CdCr_2S_4$ and $Cd_{0.97}In_{0.03}Cr_2S_4$ polycrystalline are slightly changed. The Curie temperature $T_C$, defined as the temperature corresponding to the maximum of $|dM/dT|$, is about 92.1 K for $CdCr_2S_4$ and 92.9 K for $Cd_{0.97}In_{0.03}Cr_2S_4$, respectively. After annealing, the increase of $T_c$ is less than 4 K.

Figure 1 and Figure 2 display the temperature dependence of dielectric constant $\varepsilon'$ and ac-conductivity $\sigma'$ at two different frequencies in magnetic field of 0 T and 4.5 T for the as-prepared and annealed samples. The as-prepared $CdCr_2S_4$ and $Cd_{0.97}In_{0.03}Cr_2S_4$ have similar magnitudes of $\varepsilon'$ and $\sigma'$ as the $CdCr_2S_4$ polycrystalline sample reported in Ref. 8. Both $\varepsilon'$ and $\sigma'$ increase with temperature monotonously and do not show anomaly and magnetic field dependence in the whole measured temperature range, as seen from Fig. 1(a)-(b) and Fig. 2(a)-(b). After annealing $CdCr_2S_4$ at 800 °C (Fig. 1(c)-(d)), $\varepsilon'$ displays a hump in the low temperature region and the hump temperature increases with frequency, implying a relaxor-like behavior of the dielectric property.[8] However, the magnetic field of 4.5 T still has no obvious effect on either $\varepsilon'$ or $\sigma'$.

In contrast, for the $Cd_{0.97}In_{0.03}Cr_2S_4$ sample annealed at 380 °C (Fig. 2(c)-(d)), a strong upturn of $\varepsilon'$ is clearly observed with decreasing temperature near $T_C$. $\sigma'$ has a similar temperature dependence as $\varepsilon'$. The external magnetic field of 4.5 T does not change the shape of $\varepsilon'$-T and $\sigma'$-T curves, but makes the upturn of $\varepsilon'$ and $\sigma'$ shifting towards higher temperatures. Magnetocapacitance, defined as MC = $(\varepsilon'(4.5T)-\varepsilon'(0T))/\varepsilon'(0T)$, reaches up to 290% at 1kHz and 1950% at 600 kHz (Fig. 3(a)). Both the CMC effect and the magnitudes of $\varepsilon'$ and $\sigma'$ are in agreement with that of In-doped $CdCr_2S_4$ polycrystalline sample as reported in Ref. 8. The sample annealed at 800 °C has similar temperature dependence of $\varepsilon'$ and $\sigma'$ as the sample annealed at 380 °C, while has a much higher $\varepsilon'$ and $\sigma'$, and a lower MC value ( reaches up to 145% at 1 kHz and 104% at 600 kHz , see Fig. 3(c)).

A clear feature in Fig. 1(c)-(d) and Fig 2(c)-(f) is that the appearance of CMC effect is accompanied by the field-enhanced ac conductivity. In order to investigate the correlation between the dielectric and electric transport properties, we further measured the DC-resistivity. The DC-resistivity of the as-prepared samples is too high to be measured (> 1 GΩ cm at room temperature), but is dramatically reduced after annealing. The inset of Fig. 1(d) shows the temperature dependence of DC-resistivity under 0 T and 4.5 T for the annealed $CdCr_2S_4$ sample. In accordance with the field independence of the ac-conductivity, the application of 4.5 T magnetic field has no evident influence on the DC-resistivity. Upon cooling, the DC-resistivity increases monotonously as a typical semiconductor behavior in the whole temperature range measured.

However, as seen in Fig 3(b) and Fig 3(d), the temperature dependence of DC-resistivity for the annealed $Cd_{0.97}In_{0.03}Cr_2S_4$ samples is quite different from that of $CdCr_2S_4$. For annealed $Cd_{0.97}In_{0.03}Cr_2S_4$, the zero-field resistivity first increases with decreasing temperature. After reaching up to a maximum near $T_c$, the resistivity decreases abruptly, indicting the occurrence of insulator-metal transition. Upon further cooling to the low temperature region, the resistivity increases again. Being correlated to the insulator-metal transition around $T_c$, the external magnetic field of 4.5 T makes the resistivity peak moving to a higher temperature and dramatically depresses the peak value. Magnetoresistance, defined as MR = $(\rho(0T)-\rho(4.5T))/\rho(0T)$, reaches up to about 95% for the sample annealed at 380 °C and 93% for the sample annealed at 800 °C, much higher than the value of the most investigated CMR material $FeCr_2S_4$, see the insets of Fig. 3(b) and Fig. 3(d).[15,16]

From the above-mentioned experiment results, we can know that annealing has marginal effect on the magnetism, but greatly changes the magneto-transport and magneto-dielectric behaviors. After annealing, $CdCr_2S_4$ displays a relaxor-like dielectric behavior. However, the absence of CMC strongly indicates that the relaxor-like behavior is indeed extrinsically caused by Maxwell-Wagner effect, instead by magneto-electric coupling.[17,18] The post-annealing in cadmium vapor results in heterogeneous distribution of defects in both $CdCr_2S_4$ and $Cd_{0.97}In_{0.03}Cr_2S_4$. For a Maxwell-Wagner system, CMC can appear when it is also magnetoresistive.[19] This is exactly revealed in the samples of annealed $Cd_{0.97}In_{0.03}Cr_2S_4$. The simultaneous appearance or absence of CMC and CMR in the $CdCr_2S_4$ system can thereby be well understood in the scenario of a combination of CMR and Maxwell-Wagner effect.[19]

After annealing, $CdCr_2S_4$ displays a typical semiconductor behavior, while $Cd_{0.97}In_{0.03}Cr_2S_4$ shows metal-insulator transition around $T_c$. The gigantic changes in physical properties induced by minute perturbations remind the case of $FeCr_2S_4$. For $FeCr_2S_4$, the single crystal sample displays an orbital glass state while the polycrystalline sample shows orbital ordering.[20] The frozen-in of the orbital order in single crystal $FeCr_2S_4$ is believed to be caused by chlorine impurities contaminated during the growth process, since the orbital ordering state in polycrystalline sample can be suppressed into glass state by minor element substitution.[21] For $CdCr_4S_4$ single crystals, the contamination of chlorine does exist and this could also be the reason that single crystal and polycrystalline samples display different physical properties. It should be noted that single crystal $HgCr_2S_4$ grown without fluorine flux does not show dielectric anomalies.[11] On the other hand, $CdCr_2S_4$ is conjectured to possess CMR effect which is however absent in our polycrystalline as-prepared sample.[15] We needs further investigation to understand the reason. Nevertheless, the strong sensitivity of transport properties to the detail of sample preparation implies that the $CdCr_2S_4$ is close to the boundary of insulator-metal transition, which is reflected in the electronic band structure calculations.[22,23] The calculation results of Shanthi et al. showed that $CdCr_2S_4$ corresponds to a ferromagnetic semi-metallic ground state, however, Wang et al. suggested that $CdCr_2S_4$ is a ferromagnetic semiconductor but could be modulated into a half-metal by chemical doping.[22,23]

In summary, we investigated the magnetic, dielectric and electric transport properties of $CdCr_2S_4$ and $Cd_{0.97}In_{0.03}Cr_2S_4$ polycrystalline samples before and after annealing. The observation of an one-to-one correlation between CMC and CMR implies the resistive origin of CMC in $CdCr_2S_4$ system. Since the manifestation of CMC can be entangled by the appearance of CMR, it should be handled with care to demonstrate whether the observed magneto-electric coupling is intrinsic.

**ACKNOWLEDGMENTS**

This research was financially supported by the National Key Basic Research of China Grant, Nos. 2010CB923403, and 2011CBA00111, and the National Nature Science Foundation of China Grant 11074258.

**Figure Captions:**

Figure 1. Temperature dependence of dielectric constant (upper frames) and AC conductivity (lower frames) for the as-prepared and annealed $CdCr_2S_4$ samples. Inset of (b) shows the temperature dependence of magnetization. Inset of (d) displays the temperature dependence of resistivity in magnetic field of 0 and 4.5 T for annealed $CdCr_2S_4$.

Figure 2. Temperature dependence of dielectric constant (upper frames) and AC conductivity (lower frames) for the as-prepared and annealed $Cd_{0.97}In_{0.03}Cr_2S_4$ samples. Inset of (b) shows the temperature dependence of magnetization.

Figure 3. Temperature dependence of magnetocapacitance (upper frames) and resistivity (lower frames) for annealed $Cd_{0.97}In_{0.03}Cr_2S_4$. Insets of (b) and (d) display the magnetoresistance of the samples annealed at 380   (left) and 800   (right), respectively.

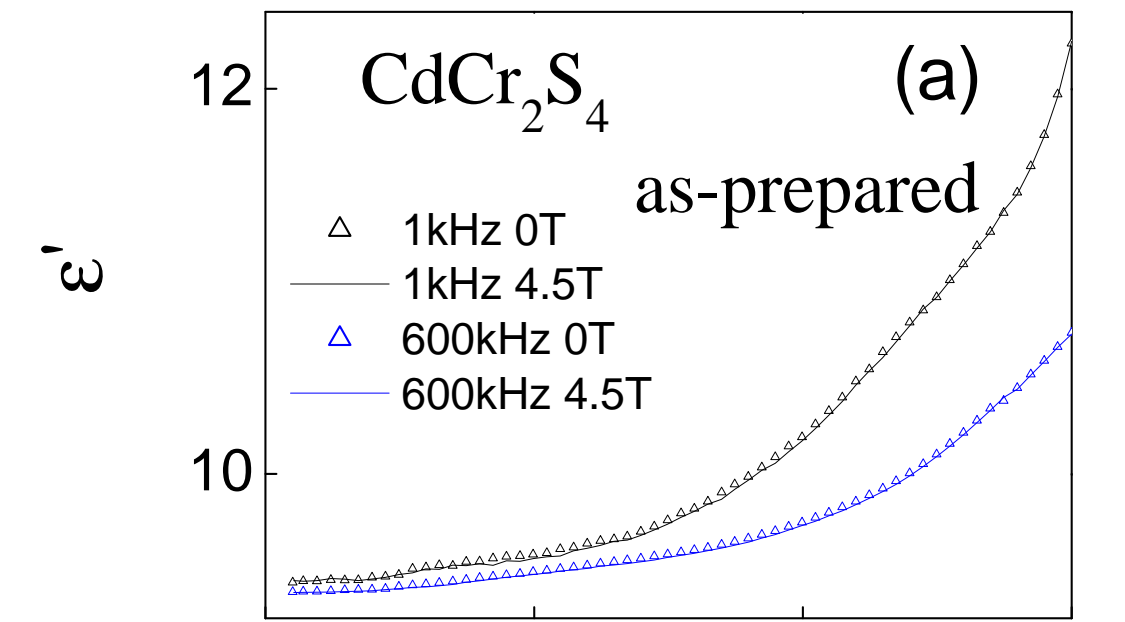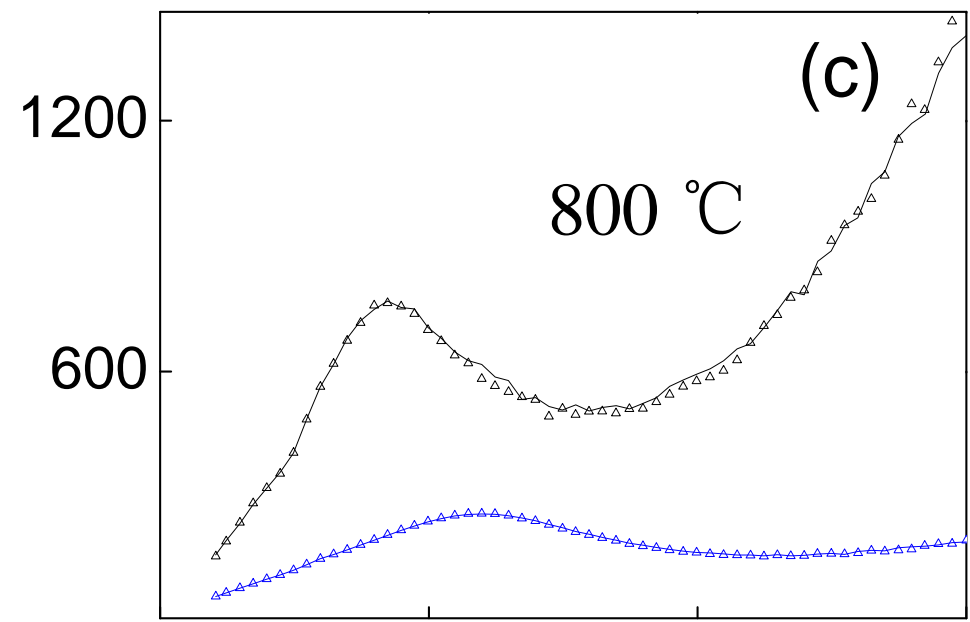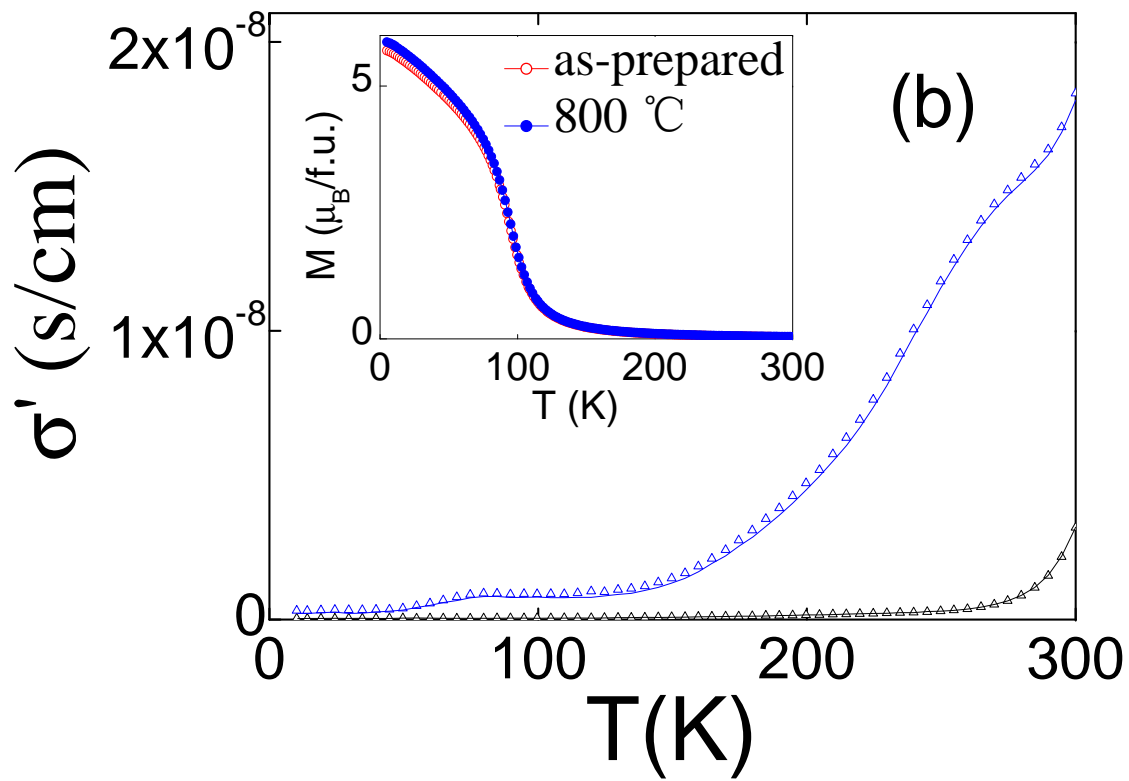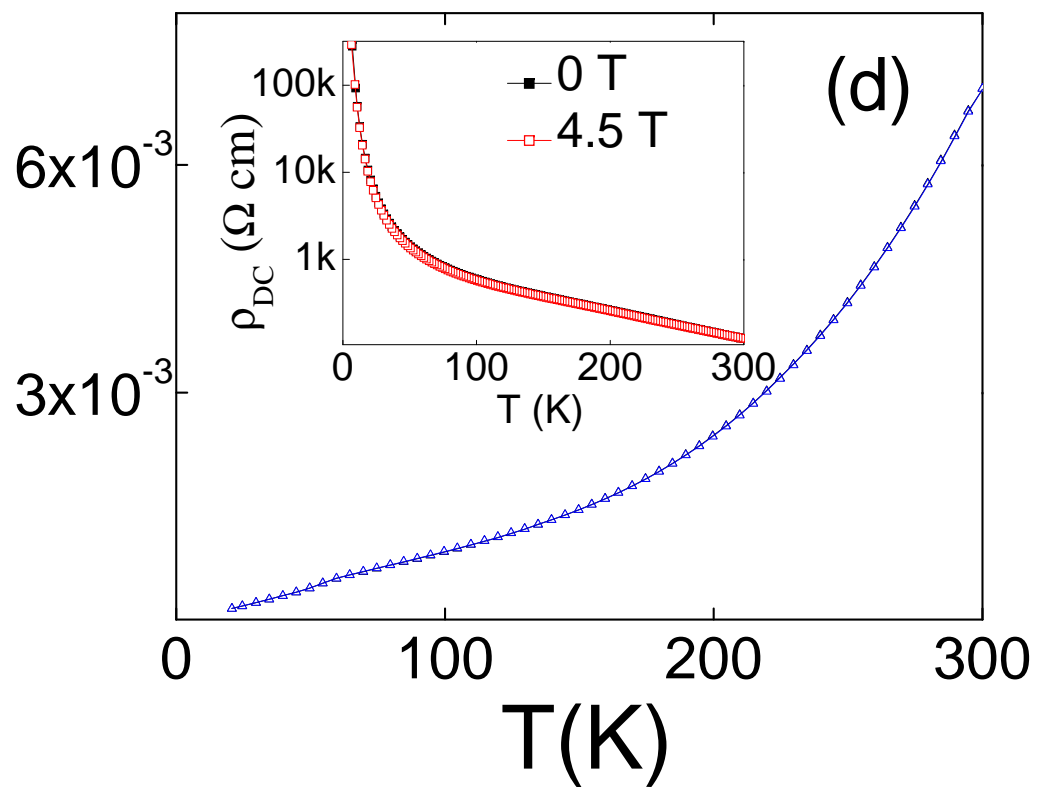

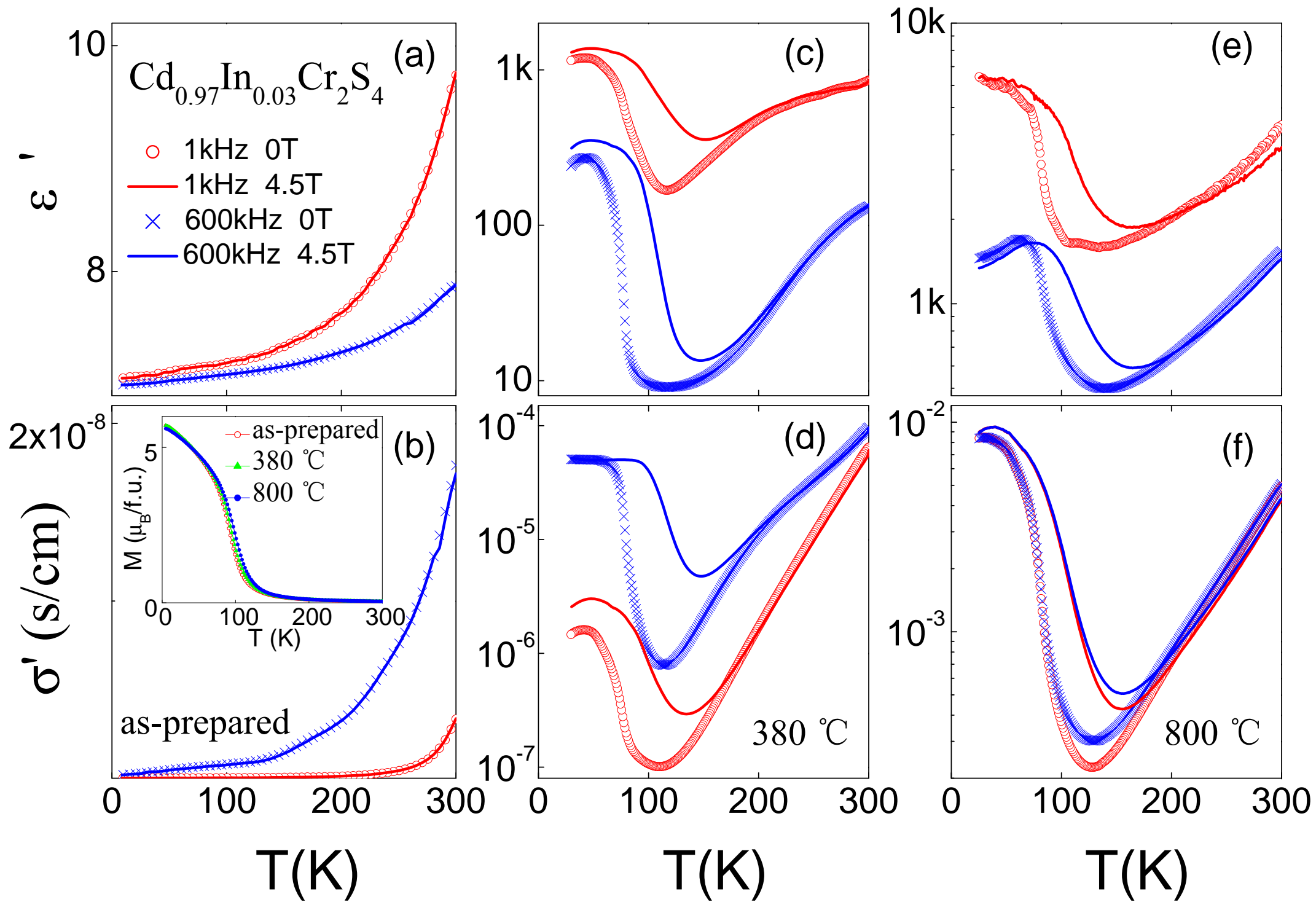

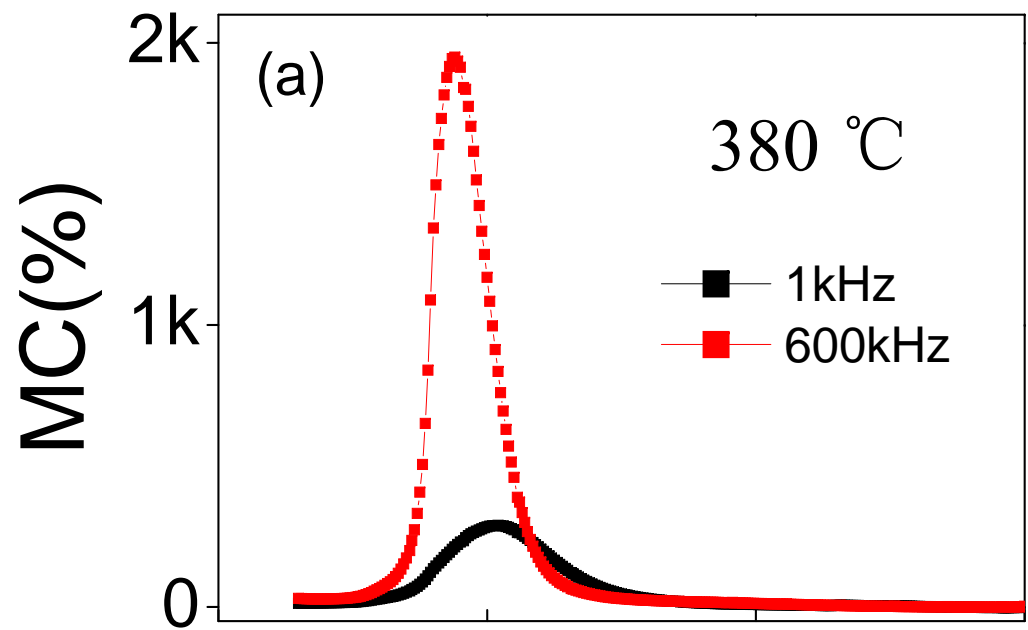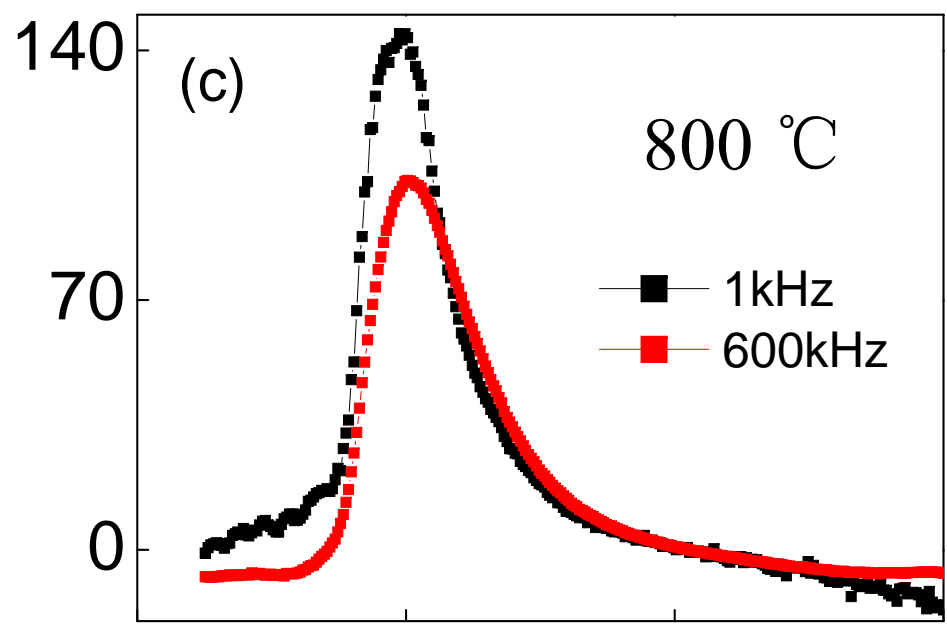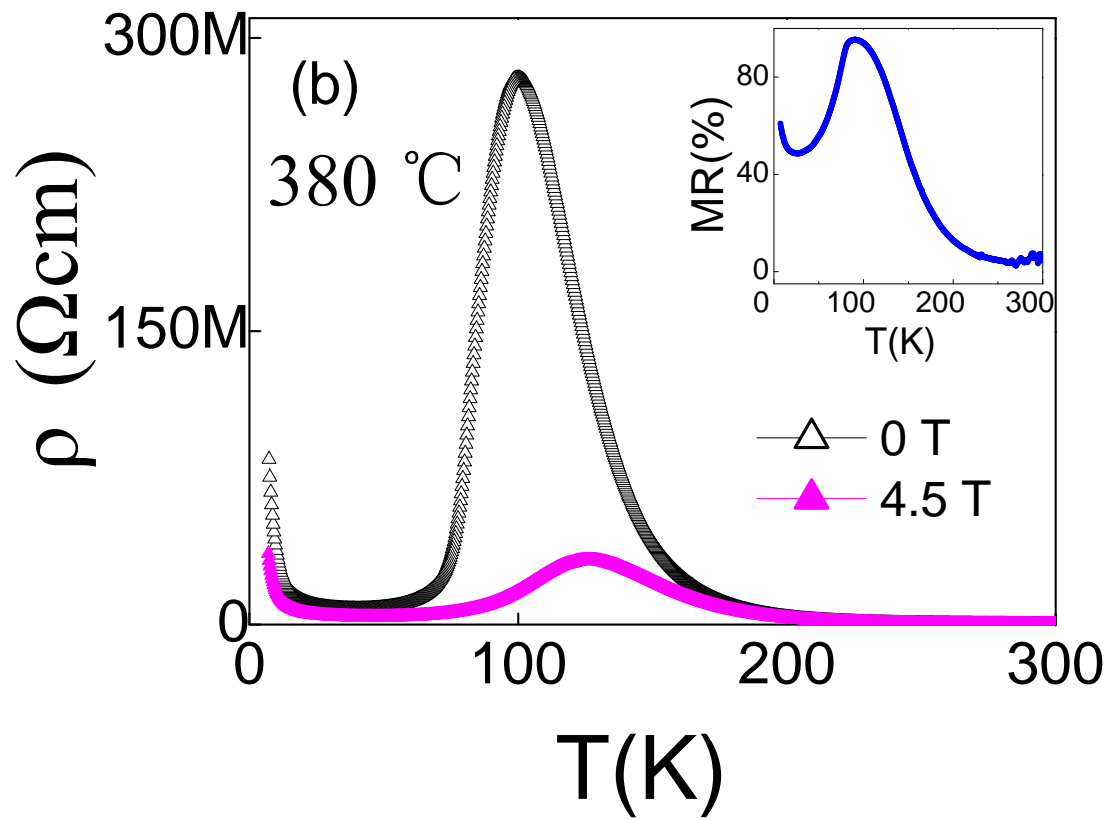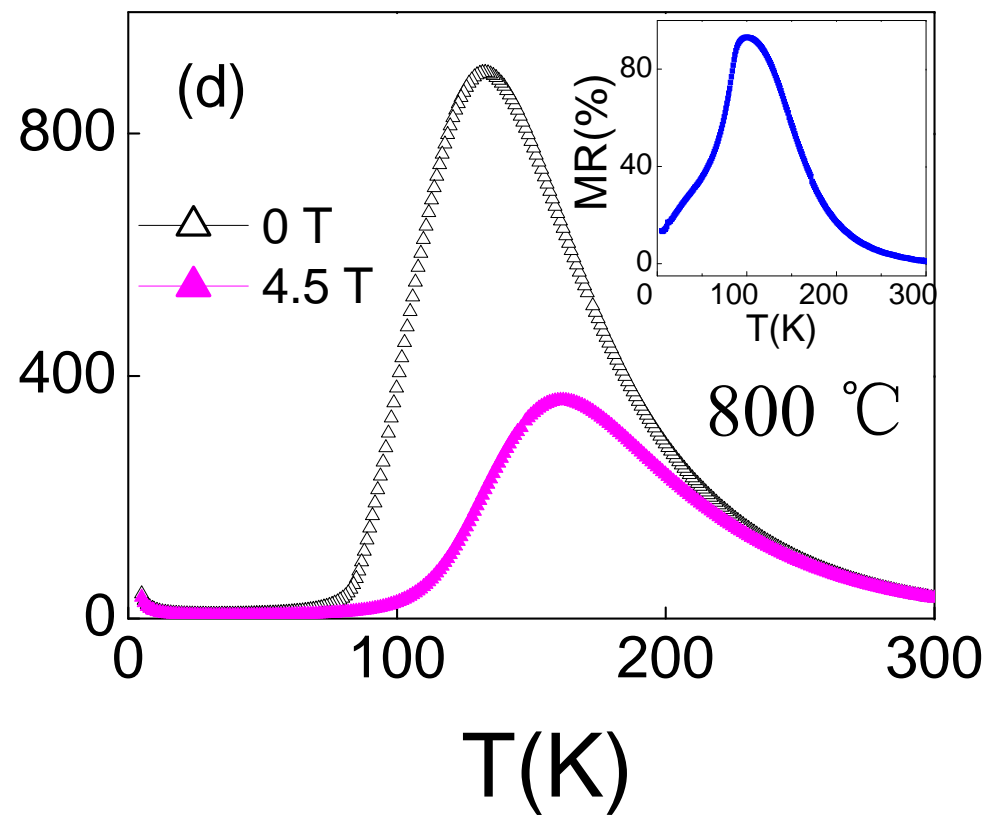